\documentclass{article}
\usepackage{spconf,amsmath,graphicx}
\usepackage{amsfonts,subfigure,cite,epsfig,etoolbox,enumitem}
\usepackage{multirow}
\usepackage{lipsum}

\title{A Global-local Attention Framework for Weakly Labelled Audio Tagging}
\name{Helin Wang$^1$, Yuexian Zou$^{1,2,*}$, Wenwu Wang$^3$ \thanks{This paper was partially supported by Shenzhen Science \& Technology Fundamental Research Programs (No: JCYJ20170817160058246,
JCYJ20180507182908274 \& JSGG20191129105421211)} \thanks{*Corresponding author: zouyx@pku.edu.cn}}
\address{
  $^1$ADSPLAB, School of ECE, Peking University, Shenzhen, China\\
  $^2$Peng Cheng Laboratory, Shenzhen, China\\
  $^3$Center for Vision, Speech and Signal Processing, University of Surrey, UK}
\begin{document}
%
\maketitle
\begin{abstract}
  Weakly labelled audio tagging aims to predict the classes of sound events within an audio clip, 
  where the onset and offset times of the sound events are not provided.
  Previous works have used the multiple instance learning (MIL) framework,
  and exploited the information of the whole audio clip by MIL pooling functions.
  However, the detailed information of sound events such as their durations may not be considered under this framework.
  To address this issue, 
  we propose a novel two-stream framework for audio tagging by exploiting the global and local information of sound events.
  The global stream aims to analyze the whole audio clip in order to capture the local clips that need to be attended using a class-wise selection module.
  These clips are then fed to the local stream to exploit the detailed information for a better decision.
  Experimental results on the AudioSet show that our proposed method 
  can significantly improve the performance of audio tagging under different baseline network architectures.
  
\end{abstract}
\section{Introduction}
\label{sec:intro}
Audio Tagging is a technique for predicting the presence or absence of sound events within an audio clip \cite{virtanen2018computational}.
The Detection and Classification of Acoustic Scenes and Events (DCASE) challenges \cite{giannoulis2013detection,stowell2015detection}
provide strongly labelled datasets for audio tagging, where the onset and offset time of sound events are annotated.
However, such annotation is time-consuming and hard to obtain, and these audio tagging datasets are relatively small.
Recently, weakly labelled audio tagging has attracted increasing interest in the audio signal processing community \cite{Serizel2018,Turpault2019},
where the datasets (\textit{e.g.} AudioSet \cite{gemmeke2017audio}) only annotate the types of sound events present in each audio clip but do not provide any timestamp information of their onset and offset.

As the duration of sound events can be very different and the overlaps of sound events often occur in an audio clip,
audio tagging with weakly labelled data is a challenging problem.
A popular approach for this problem is based on multiple instance learning (MIL) \cite{amores2013multiple,tseng2017multiple}.
In MIL, the input sequence is treated as a bag and split into the set of instances, 
where multiple instances in the same bag share the same labels.
There are two main MIL strategies, \textit{i.e.} instance-level approach \cite{wang2019comparison} and embedding-level approach \cite{kao2020comparison,lin2020specialized}.
The embedding-level approach integrates the instance-level feature representations into a bag-level contextual representation and then directly carries out bag-level classification,
which shows better performance than the instance-level approach \cite{wang2018revisiting}.
The methods for aggregating the information from the instances play an important part in the MIL frameworks.
The default choices are global max pooling (GMP) \cite{oquab2015object} and global average pooling (GAP) \cite{zhou2016learning},
but they are often less flexible for adapting to practical applications and may lose detailed information relevant to acoustic events. 
For example, GMP cannot capture the information of the long-duration event \textit{keyboard typing} well, 
while the short-duration event \textit{mouse click} may be ignored by GAP.
More recently, attention mechanisms have been employed to detect the occurence of sound events \cite{kong2019weakly,yu2018multi,ilse2018attention,wang2020environmental}
and have achieved promising results.
However, these methods attempt to make a decision on the whole audio clips and are limited in capturing the detailed information of the acoustic events.

\begin{figure*}[t]
  \centerline{\includegraphics[width=6.5in]{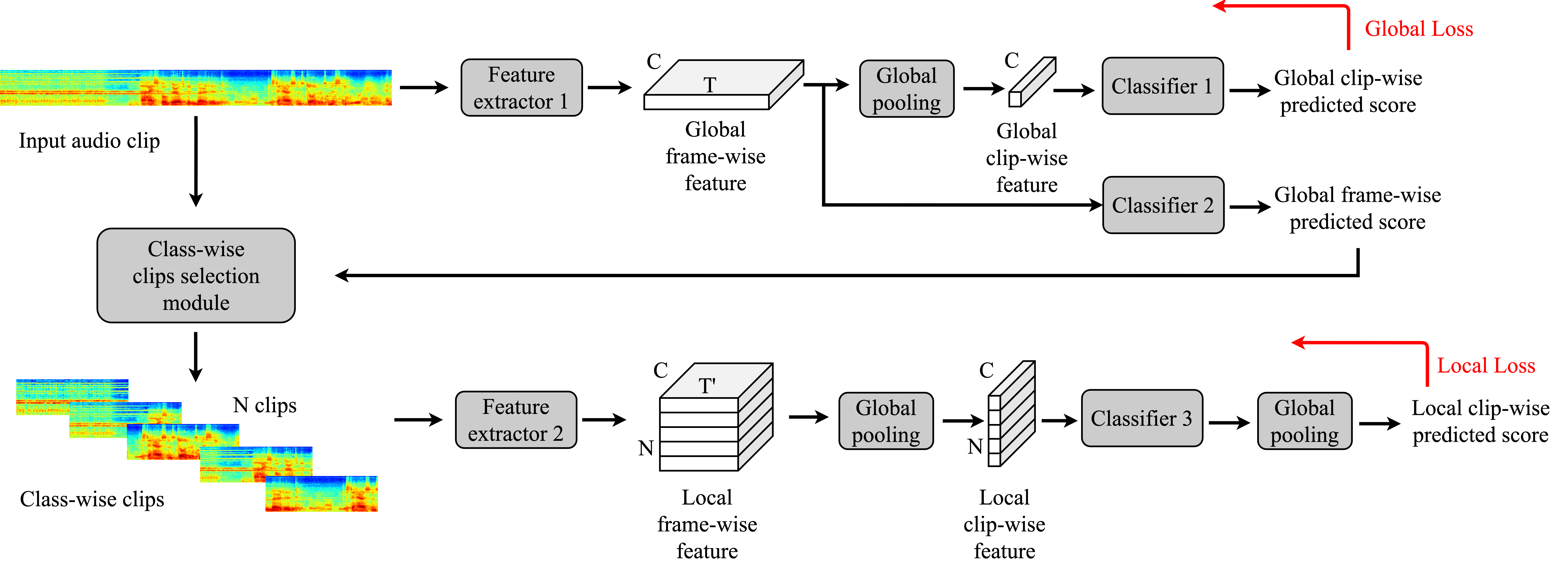}}
  \caption{Overall architecture of our two-stream framework for weakly labelled audio tagging (GL-AT).
  }  
  \label{fig1}
  \end{figure*}

To address this issue,
in this paper,
we propose a novel global-local attention (GL-AT) framework for weakly labelled audio tagging where the global and local information are successively modelled in the audio clip.
Our method is inspired by the behaviors of human annotators for an audio dataset \cite{gemmeke2017audio}.
At first, the annotators may glimpse over an audio clip roughly, and determine some possible categories and their temporal regions.
Then, these possible regions guide them to make refined decisions on specific categories following a region-by-region inspection.
Thus, in a similar fashion, we can solve the audio tagging with a two-stream framework, including a global stream and a local stream.
More specifically,
the global stream takes an audio clip as the input to a deep neural network 
and learns global representations supervised by the weak labels.
Several class-wise attentional sub-clips are then selected according to the global representations,
and fed to another neural network to learn the local representations.
The final class distributions are obtained by aggregating the predicted global class distributions and local class distributions.
The optimization of the local stream is influenced by the global stream because the sub-clips are selected according to the prediction results of the global stream. 
In addition, the local stream improves the optimization of the global stream where sub-clip selection and
classification are both performed.

The contributions of this paper can be summarized into three aspects.
Firstly, we present a two-stream global-local attention framework that can efficiently recognize sound events within an audio clip
with weakly labelled data. 
Secondly, we propose an effective class-wise sub-clip selection module which
can dynamically generate several attentional sub-clips with low complexity and high diversity.
Thirdly, experimental results show that our proposed framework can significantly improve the performance of 
AudioSet tagging, and can be used in different baselines.


\section{Proposed Method}
\label{method}
In this section,
we firstly present the two-stream framework for weakly labelled audio tagging (\textit{i.e.} GL-AT). 
Then, a class-wise clips selection module is presented, 
which bridges the gap between the global and local streams. 

\subsection{GL-AT}
\noindent
\textbf{Global Stream.} The overall framework of our proposed framework is shown in Fig.~\ref{fig1}.
Given an input audio clip $\boldsymbol{A} \in \mathbb{R}^{T_{0} \times S}$,
where $T_{0}$ and $S$ are the duration (\textit{e.g.} $10$s) and the sampling rate, respectively.
Let’s denote its corresponding label as $\boldsymbol{y}\in \mathbb{R}^{L}$,
where $y^{i}=\{0,1\}$ denotes whether label $i$ appears or not and $L$ denotes the number of labels.

The feature extractor $\mathcal{F}(\cdot)$ is firstly applied, 
which can be either convolutional neural networks (CNNs) \cite{hershey2017cnn,choi2016automatic,xu2018large,kong2020panns,Ford2019} or convolutional recurrent neural networks (CRNNs) \cite{wang2019comparison}.
We assume that $\boldsymbol{M} = \mathcal{F}\left(\boldsymbol{A};\theta_{\mathcal{F}}\right)$ is the global frame-wise feature
after the feature extractor, where $\theta_{\mathcal{F}}$ denotes the parameters of the feature extractor
and $\boldsymbol{M} \in \mathbb{R}^{T \times C}$.
Here, $T$ and $C$ denote the number of output frames and the dimension of the features for each frame, respectively.
Then a global pooling function $\mathcal{P}\left(\cdot\right)$ is applied to obtain the global clip-wise feature $\boldsymbol{M^{\prime}} \in \mathbb{R}^{1 \times C}$.
Following \cite{kong2020panns}, both maximum and average operations are used for global pooling.
In order to get the prediction score $\hat{\boldsymbol{y}}\in \mathbb{R}^{L}$, 
a classifier $\mathcal{C}(\cdot)$ containing two fully-connected layers is applied \cite{kong2020panns}.
\begin{equation}
  \hat{\boldsymbol{y}} = \mathcal{C}\left(\boldsymbol{M^{\prime}};\theta_{\mathcal{C}}\right)
  \label{eq1}
\end{equation}
where $\theta_{\mathcal{C}}$ denotes the parameters of the classifier.
We then use a sigmoid function $\sigma\left(\cdot\right)$ to turn $\hat{\boldsymbol{y}}$ into
a range $\left[0,1\right]$, and obtain the global clip-wise prediction score $\hat{\boldsymbol{y}}_{\boldsymbol{g}}\in \mathbb{R}^{L}$.
\begin{equation}
  \hat{\boldsymbol{y}}_{\boldsymbol{g}} = \frac{1}{1 + exp\left(-\hat{\boldsymbol{y}}\right)}
  \label{eq2}
\end{equation}

\noindent
\textbf{Local Stream.} Let $\left\{\boldsymbol{A}_{1}, \boldsymbol{A}_{2}, \cdots, \boldsymbol{A}_{N}\right\}$
be a set of $N$ local clips selected from the input audio clip $\boldsymbol{A}$.
These local clips have the same duration (but are shorter than $\boldsymbol{A}$), 
and are fed to another feature extractor which has the same structure as the global stream.
Then the local prediction scores $\left\{\hat{\boldsymbol{y}}_{1}, \hat{\boldsymbol{y}}_{2}, \cdots, \hat{\boldsymbol{y}}_{N}\right\}$ are obtained using (\ref{eq1}) and (\ref{eq2}).
Finally, these local predicted scores are aggregated by the global pooling function:
\begin{equation}
  \hat{\boldsymbol{y}}_{\boldsymbol{l}}=\mathcal{P} \left(\hat{\boldsymbol{y}}_{1}, \hat{\boldsymbol{y}}_{2}, \cdots, \hat{\boldsymbol{y}}_{N}\right)
  \label{eq3}
\end{equation}
where $\hat{\boldsymbol{y}}_{\boldsymbol{l}} \in \mathbb{R}^{L}$ is the local clip-wise prediction score.
Note that this two-stream framework can be trained end-to-end and transferred easily to different feature extractor networks.
During the training stage, these two streams are jointly trained.
At the inference stage, we fuse the predictions from the global stream ($\hat{\boldsymbol{y}}_{\boldsymbol{g}}$)
and the local stream ($\hat{\boldsymbol{y}}_{\boldsymbol{l}}$) with the global pooling function
to generate the final prediction score of the audio.

\noindent
\textbf{Two-stream Learning.}
Given a training dataset $\left\{\boldsymbol{A}^{i}, \boldsymbol{y}_{i}\right\}_{i=1}^{D}$,
where $D$ denotes the number of training examples.
$\boldsymbol{A}^{i}$ is the $i$-th audio clip and $\boldsymbol{y}_{i}$ represents its corresponding labels.
The overall loss function of our two-stream learning is formulated as the sum of two streams,
\begin{equation}
  \boldsymbol{\mathcal{L}}=\boldsymbol{\mathcal{L}_{g}}+\boldsymbol{\mathcal{L}_{l}}
  \label{eq4}
\end{equation}
where $\boldsymbol{\mathcal{L}_{g}}$ and $\boldsymbol{\mathcal{L}_{l}}$ represent the global and the local loss, respectively. 
Specifically, the binary cross entropy loss is applied for both streams,
\begin{equation}
  \boldsymbol{\mathcal{L}_{g}}=\sum_{i=1}^{D} \sum_{j=1}^{L} y_{i}^{j} \log \left(\hat{y}_{g i}^{j}\right)+\left(1-y_{i}^{j}\right) \log \left(1-\hat{y}_{g i}^{j}\right)
  \label{eq5}
\end{equation}
\begin{equation}
  \boldsymbol{\mathcal{L}_{l}}=\sum_{i=1}^{D} \sum_{j=1}^{L} y_{i}^{j} \log \left(\hat{y}_{l i}^{j}\right)+\left(1-y_{i}^{j}\right) \log \left(1-\hat{y}_{l i}^{j}\right)
  \label{eq6}
\end{equation}
where $\hat{y}_{g i}^{j}$ and $\hat{y}_{l i}^{j}$ are the prediction scores of the $j$-th category of the 
$i$-th audio clip from the global stream and local stream, respectively.
The Adam \cite{kingma2014adam} is employed as the optimizer.

\subsection{From Global to Local}
\label{gtl}
Potential audio clips are not available in weak labels, and in this paper,
we propose a simple but efficient method to dynamically generate candidate audio clips following two basic principles.
On the one hand, the diversity of candidate clips should be as high as possible to cover all possible sound events within an audio clip.
On the other hand, the number of candidate clips should be as small as possible to reduce computational complexity and storage space.

In order to generate the candidate clips, we firstly calculate the class-wise activation.
A classifier is directly applied to the global frame-wise feature $\boldsymbol{M}$, 
and the global frame-wise predicted score $\boldsymbol{S}_{g} \in \mathbb{R}^{T \times L}$ can be obtained using (\ref{eq1}) and (\ref{eq2}),
where $T$ denotes the number of frames within an audio clip.
The class-wise activation of the $i$-th category is denoted as $S_{g}^{i} \in \mathbb{R}^{T}$,
which indicates the importance of the frames leading to the classification of an audio clip to class $i$.

Such class-wise activation $\boldsymbol{S}_{g}$ is discriminative among different categories,
and we can employ the activation to localize the potential candidate clips.
However, there are often a large number of categories (\textit{e.g.} $527$ categories in AudioSet),
and only a small number of categories (less than $10$) appear in an audio clip.
If the activation of all the categories is used, the generated clips are too many to be computationally efficient.
To address this issue, we sort the predicted score $\boldsymbol{S}_{g}$ in a descending order 
and select the top $N$ class-wise attentional activation (denoted as $\left\{S_{g}^{i}\right\}_{i=1}^{N}$).
$N$ is a hyperparameter and the choice of $N$ is discussed in our experiments.

The value of $S_{g}^{i}(t)$ represents the probability 
that the sub-clip belongs to the $i$-th category at timestamp $t$.
In order to localize the clips of interest with low computational complexity,
we employ a temporal window of size $\tau$ to select the candidate clips.
For each $S_{g}^{i}(t)$, the frame with the maximum activation (denoted as $m$) is set as the medium frame,
and the range of the candidate clip is set as $\left[m-\tau/2,m+\tau/2\right]$.
If the maximum boundary of the candidate clip is over the duration of the audio clip,
the range is re-weighted to $\left[T_{0}-\tau,T_{0}\right]$, where $T_{0}$ denotes the duration of the audio clip.
In addition, if the minimum boundary is less than $0$,
we re-weight the range to $\left[0,\tau\right]$,
where $\tau$ is a hyperparameter and discussed in our experiments.

\section{Experiments}
\label{exp}
In this section, we report the experimental results and comparisons that demonstrate the effectiveness
of the proposed method. 
In addition, ablation studies are carried out to show the contribution of the crucial components.

\subsection{Experimental Setups}
\noindent
\textbf{Dataset.} Audioset \cite{gemmeke2017audio} is used in our experiments,
which is a large-scale dataset with over $2$ million $10$-second audio clips from YouTube videos, with a total of $527$ categories.
The same dataset divisions and pre-processing approaches (\textit{e.g.} re-sampling and data-balancing) are applied as \cite{kong2020panns}.

\noindent
\textbf{Metrics.} Mean average precision (mAP), mean area under the curve (mAUC) and d-prime are used as our evaluation metrics,
which are the most commonly used metrics for audio tagging.
These metrics are calculated on individual categories and then averaged across all categories.

\noindent
\textbf{Implementation Details.} We compare the proposed method with the recent state-of-the-arts
including TALNet \cite{wang2019comparison}, CNN10 \cite{kong2020panns}, ResNet38 \cite{kong2020panns} and AT-SCA \cite{hong2020weakly}.
Specifically, all these models are applied as the global stream of our framework,
and the local stream uses the same feature extractor and classifier as in the global stream.
Thus, we obtain the results of the models with our two-stream learning framework,
which are TALNet\textcircled{+}GL-AT, CNN10\textcircled{+}GL-AT, ResNet38\textcircled{+}GL-AT and AT-SCA\textcircled{+}GL-AT.
Batch size is set as $32$ and all networks are trained with $800$k iterations in total.
Unless otherwise stated, we set the hyperparameters $N$ as $5$ and $\tau$ as $3$s in our experiments. 

\subsection{Comparsions with the State-of-the-Arts}
Table~\ref{t1} demonstrates the performance of our proposed method (GT-AL\footnote{https://github.com/WangHelin1997/GL-AT}) and other state-of-the-art methods on the Audioset.
The results indicate that the proposed GT-AL can significantly improve the performance of all the compared methods,
which confirms the effectiveness of our two-stream learning framework.
Among them, TALNet \cite{wang2019comparison} is a CRNN-based model, and CNN10 \cite{kong2020panns} and ResNet38 \cite{kong2020panns} are 
CNN-based models.
AT-SCA \cite{hong2020weakly} applies the spatial and channel-wise attention to detect sound events by learning what and where to attend in the signal.
However, all these methods attempted to make a decision once and aggregated the global information by the MIL pooling functions.
The aggregation often cannot make good use of the detailed information because of the variability of different sound events.
By comparison, our method captures the discriminative local information from the global information and then checks attentionally the local information for a better decision,
which provides a more efficient way to exploit the weak labels.

\begin{table}[t]
  \begin{center}
  \caption{Accuracy comparisons of our method and state-of-the-arts on the AudioSet.} \label{t1}    
  \begin{tabular}{|c|c|c|c|}
      \hline
      \textbf{Method} & \textbf{mAP} & \textbf{mAUC} & \textbf{d-prime}\\
      \hline
      TAL Net (2019) \cite{wang2019comparison} & 0.362 & 0.965 & 2.554\\
      TAL Net$^{\mathrm{\ast}}$ & 0.368 & 0.967 & 2.600\\
      \textbf{TALNet}\textcircled{+}\textbf{GL-AT (ours)} & \textbf{0.401} & \textbf{0.970} & \textbf{2.659}\\
      \hline
      CNN10 (2019) \cite{kong2020panns} & 0.380 & 0.971 & 2.678\\
      CNN10$^{\mathrm{\ast}}$ & 0.382 & 0.969 & 2.664\\
      \textbf{CNN10}\textcircled{+}\textbf{GL-AT (ours)} & \textbf{0.408} & \textbf{0.974} & \textbf{2.742}\\
      \hline
      ResNet38 (2019) \cite{kong2020panns} & 0.434 & 0.974 & 2.737\\
      ResNet38$^{\mathrm{\ast}}$ & 0.429 & 0.974 & 2.713\\
      \textbf{ResNet38}\textcircled{+}\textbf{GL-AT (ours)} & \textbf{0.438} & \textbf{0.975} & \textbf{2.774}\\
      \hline
      AT-SCA (2020) \cite{hong2020weakly} & 0.390 & 0.970 & 2.652\\
      AT-SCA$^{\mathrm{\ast}}$ & 0.392 & 0.969 & 2.658\\
      \textbf{AT-SCA}\textcircled{+}\textbf{GL-AT (ours)} & \textbf{0.413} & \textbf{0.971} & \textbf{2.677}\\
      \hline
      \multicolumn{4}{p{230pt}}{ $^{\mathrm{\ast}}$The listed results of TAL Net, CNN10, ResNet38 and AT-SCA are reproduced,
      and all the experimental setups are the same as the original papers \cite{wang2019comparison,kong2020panns,hong2020weakly}.}\\ 
\end{tabular}
\label{t1}
\end{center}
\end{table}

\begin{table}[t]
  \begin{center}
  \caption{ Ablative study of two streams in CNN10\textcircled{+}GL-AT} \label{t2}    
  \begin{tabular}{|c|cc|c|c|c|}
      \hline
      \textbf{Method} & \textbf{Global} & \textbf{Local} & \textbf{mAP} & \textbf{mAUC}\\
      \hline
      CNN10 & $\surd$ & & 0.382 & 0.969\\
      \hline
      \multirow{3}{*}{CNN10\textcircled{+}GL-AT} & $\surd$ & & 0.389 & 0.970\\
      &  & $\surd$ & 0.400 & 0.972\\
      & $\surd$ & $\surd$ & 0.408 & 0.974\\
      \hline
\end{tabular}
\label{t2}
\end{center}
\end{table}

\subsection{Ablation Study}
In order to explore the effectiveness of the two streams, 
we jointly train the global and local streams in GL-AT, 
and during the inference stage, the influence of each stream is demonstrated in Table~\ref{t2}.
Thanks to the joint training strategy, 
GL-AT outperforms the baseline method (CNN10) with only the global stream.
This is because the class-wise attentional clips selection module connects the two streams,
and the optimization of the global stream is influenced by the local stream, 
which leads to better robustness.
In addition, we can see that using the local stream alone performs better than only using the global stream,
for the reason that the local stream is able to focus on the detailed information of the audio.
Nonetheless, the global stream plays an important role in guiding the learning of the local stream,
and employing both global and local streams achieves the best results in our experiments.

\begin{figure}[t] 
  \centering
  \subfigure[]{\label{f1}
  \includegraphics[width=1.6in]{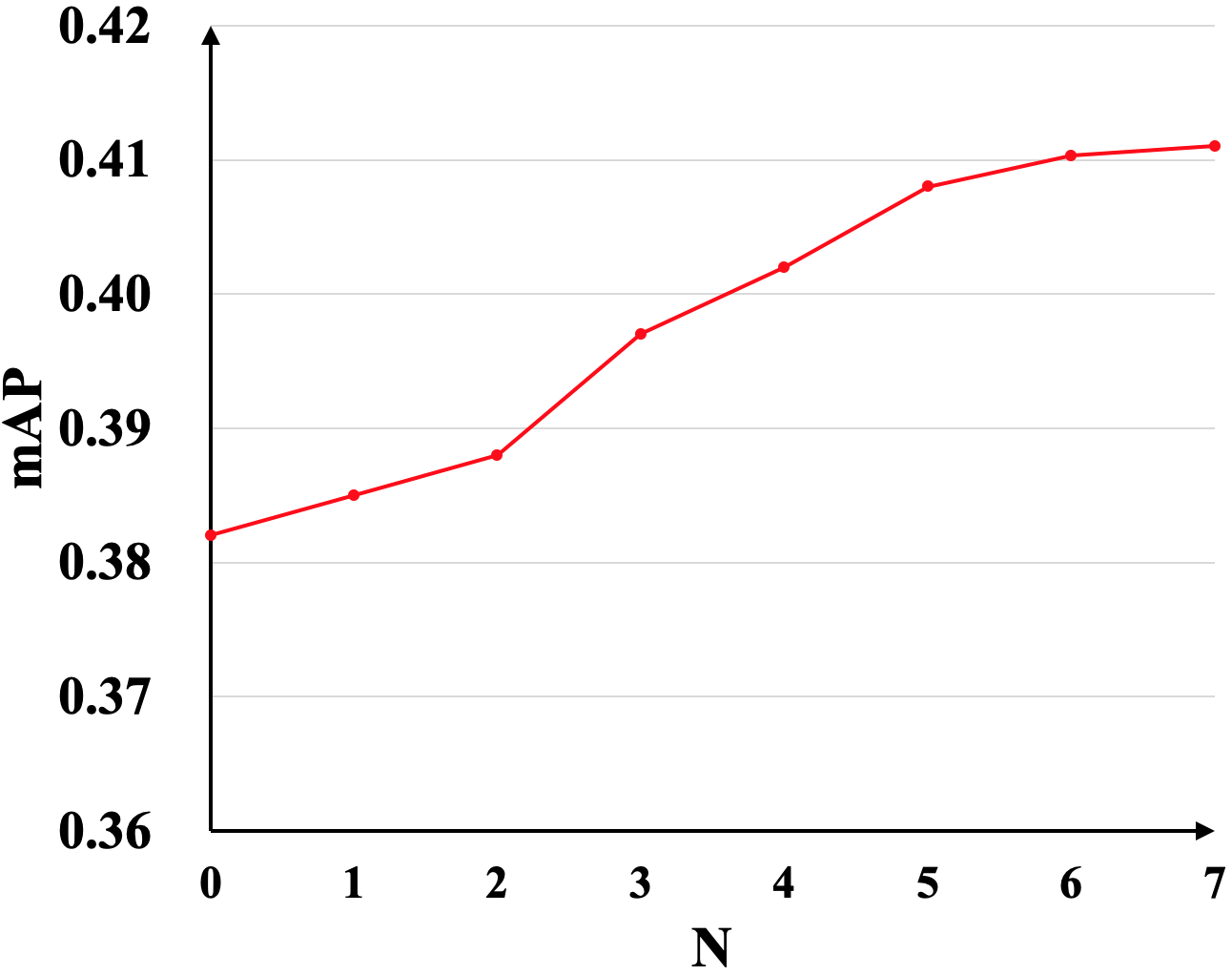}}
  \hspace{-2mm}
  \subfigure[]{\label{f2}
  \includegraphics[width=1.6in]{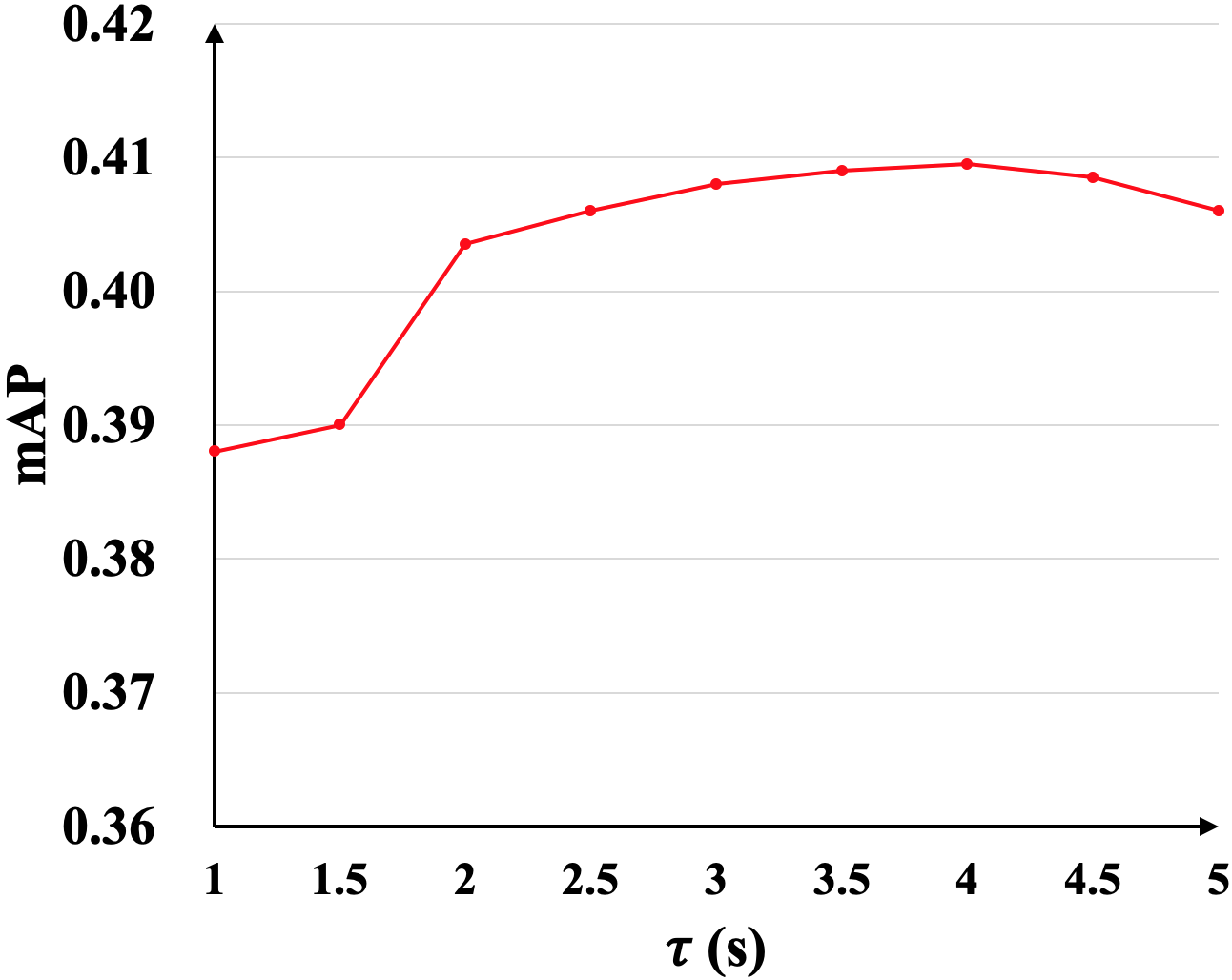}}
  \caption{ Accuracy comparisons of CNN10\textcircled{+}GL-AT with different values of $N$ and $\tau$ (metric: mAP)}
  \label{fig2}
\end{figure}

Furthermore, we exploit the influence of the number of the local clips ($N$) and the duration of the local clips ($\tau$).
As shown is Fig.~\ref{f1}, 
the mAP performance shows an upward trend with the gradual increase in $N$.
This means that it is useful to improve the audio tagging performance with more local clips.
However, more clips will increase the computational cost because all the selected clips are fed to the networks. 
We can see that the performance tends to be stable when $N$ is set to $\left\{5,6,7\right\}$,
and we set $N$ to $5$ for a balance of accuracy and complexity.
In addition, we test different $\tau$ values,
and report the results in Fig.~\ref{f2}.
As the values of $\tau$ increase, the accuracy is boosted and then drops, 
which achieves high accuracy when $\tau$ is $\left[3,4\right]$ (similar to \cite{salamon2014dataset}).
We argue that the length (\textit{i.e.} about $3s$) is the duration of most sound events.
If the duration of the clips is too short (\textit{e.g.} $1s$), the complete information of a sound event cannot be captured.
While on the other hand, if the duration is too long (\textit{e.g.} over $4s$), the detailed information will be compromised.
The extreme is that when the duration is $10s$ (the whole length of the audio clip), the local stream works as the global stream,
which does not capture the detailed information intended.

\section{Conclusions}
\label{conc}
We have presented a two-stream framework to take advantage of the global and local information of audio,
which resembles the multi-task learning principle.
Experimental results on the Audioset show that our method can boost the performance of different state-of-the-art methods in audio tagging.


\small
\bibliographystyle{IEEEtran}
\bibliography{strings,refs}

\end{document}